\documentclass[aps,prb,twocolumn,epsf,floatfix]{revtex4}
\usepackage{graphicx}
\usepackage{amsmath}
\newcommand{\rr}{{\bf r}}

\newcommand{\kk}{{\bf k}}

\newcommand{\qq}{{\bf q}}

\newcommand{\GG}{{\bf G}}

\begin{document}
\title{Theory of Spin-Charge Coupled Transport in a Two-Dimensional 
Electron Gas with the Rashba Spin-Orbit Interactions}
\author{A.A. Burkov}
\affiliation{\small Department of Physics, University of California,
Santa Barbara, CA 93106} 
\author{Alvaro S. N\'u\~nez and A.H. MacDonald}
\affiliation{\small Department of Physics, The University of Texas at 
Austin, Austin, TX 78712}
\date{\today}
\begin{abstract}
We use microscopic linear response theory to derive a set of equations that provide a
complete description of coupled spin and charge diffusive transport in 
a two-dimensional electron gas (2DEG) with the Rashba spin-orbit (SO) 
interaction. These equations capture a number of interrelated effects 
including spin accumulation and diffusion, 
Dyakonov-Perel spin relaxation, magnetoelectric, and spin-galvanic effects.
They can be used under very general circumstances to model transport 
experiments 
in 2DEG systems that involve either electrical or optical spin injection.  
We comment on the relationship between these equations and the exact 
spin and charge density operator equations of motion.
As an example of the application of our equations, we consider 
a simple electrical spin injection experiment and show that a 
voltage will develop 
between two ferromagnetic contacts if a spin-polarized current is injected 
into a 2DEG, that depends on the relative magnetization orientation of the
contacts. 
This voltage is present even when the separation between the contacts is 
larger than the spin diffusion length. 
\end{abstract}

\maketitle

\section{Introduction}
\label{sec:1}
Spintronics is an active field which studies processes that
manipulate and probe the electronic spin degree-of-freedom,
with the goal of identifying effects that can augment the orbital control
and measurement procedures used in traditional electronics.\cite{Awschalom02}
Spin-related transport effects in ferromagnetic metals are already used in
current technology to provide the robust and responsive magnetic 
field sensors required by magnetic information storage systems.  
Hopes that spin-related transport effects might 
play a greater role in future information processing and storage technologies have 
motivated a growing body of research on the creation of spin-polarized carrier distribution 
in semiconductors, either optically \cite{Awschalom98} or by injection 
from other magnetic systems.\cite{Molenkamp99,Johnson,Awschalom99}  
Semiconductor quantum well electron gas systems are especially promising for 
spintronics because their intrinsic spin-orbit (SO) interactions are weak, 
implying long spin memory times, and because the Rashba SO interaction,\cite{Rashba84}
which enables electrical control of spin, can be tuned over a wide range by
applying growth-direction gate potentials.\cite{Nitta} 

Although the study of spin dynamics in semiconductors in the presence of SO 
interactions was initiated a rather long time 
ago,\cite{Perel71,Rashba84,Dyak86} it continues to pose interesting and 
challenging problems. 
The Rashba SO interaction \cite{Rashba84} has received special attention, 
in part
because of a proposal by Datta and Das \cite{Das90} that it could be exploited in a 
{\em spin transistor}---a device in which currents are modulated by using a 
gate to alter the Rashba interaction strength.
Some interesting refinements of the original idea have appeared
in recent literature.\cite{Zulicke02}
More generally, there has recently been substantial theoretical work on 
spin-dependent transport in a 2DEG with the Rashba and other types 
of SO interactions; see,  
e.g. Refs.[\onlinecite{Levitov85,Malshukov96,Hirsch99,Magarill01,
Raimondi01,Froltsov01,Saikin03,Pershin03,Molenkamp01,Bruno02,Gridnev02,Pala02,Inoue03,Halperin03,Inoue04}].
Diffusion equations valid for weak SO interactions, which capture effects
of the Rashba spin precession beyond the Dyakonov-Perel theory, have been 
derived
and studied.\cite{Malshukov96,Froltsov01,Saikin03,Pershin03}

In this paper we derive a set of equations that provide a
complete description of coupled spin and charge diffusive transport in
a 2DEG with the Rashba SO interactions.
These equations capture a number of interrelated effects including spin accumulation 
and diffusion, Dyakonov-Perel spin relaxation and magnetoelectric and spin-galvanic effects. \cite{Ganichev02}  
This unified description is essential, since spin 
transport is most easily detected in practice through the spin accumulation it
induces at the edges of the sample.\cite{vanWees}  
A complete understanding of the interrelated spin accumulation 
and magnetoelectric effects in a given experimental situation can be 
obtained by solving the equations derived below.
Our derivation is based on a microscopic evaluation of the disorder-averaged
density-matrix response function, followed by  
an analysis of its long-wavelength, low frequency limit.  
We apply our equations to a simple model of electrical spin injection into 
a 2DEG from ferromagnetic spin-polarized contacts, placed on top of the 
2DEG.  
We find that a voltage develops between  
ferromagnetic contacts when a spin polarized current is injected into the 
2DEG, that depends on the relative magnetization 
orientation of the contacts.
Unlike all other known magnetoresistive effects in spintronics, 
this voltage drop is present even when the distance between the electrodes 
exceeds the spin-diffusion length.
  
The paper is organized as follows. 
In section \ref{sec:2} we outline the density matrix response function 
formalism that we use to derive our spin and charge transport equations.
In section \ref{sec:3} we derive transport equations from the 
low frequency, long-wavelength limit of the density matrix response function.
Section \ref{sec:4} is devoted to a discussion of the relationship between 
formally exact equations of motion for charge and spin-density operators 
in systems with Rashba spin-orbit interactions and arbitrary 
scalar disorder potentials and the coarse-grained 
dynamics predicted by our diffusive transport equations. 
Section \ref{sec:5} comments on the physical content of these equations and 
discusses an application to the case of electrical spin-injection from spin
polarized contacts.  Finally, in Section \ref{sec:6} we briefly summarize our 
findings. 

\section{Density-Matrix Response Function}
\label{sec:2} 
Our analysis of coupled 
spin and charge transport in a semiconductor 2DEG 
system uses a model of noninteracting electrons described by an 
effective-mass Hamiltonian, moving in a random short-range spin-independent
impurity potential. 
Because of the externally controllable inversion-asymmetry of the quantum
well confining potential, 
electrons experience a tunable SO interaction that we assume to be of 
the Rashba type.\cite{Rashba84}
The system is therefore described by a single-particle Hamiltonian $H=H_0+H_i$ where
\begin{equation}
\label{eq:1}
H_0=\sum_{\kk \sigma \sigma'}\left(\frac{\kk^2}
{2m}-\mu +\lambda \hat z \cdot [\boldsymbol{\tau}_{\sigma\sigma'} \times \kk]
\right) c^{\dag}_{\kk \sigma} c^{\vphantom \dag}_{\kk \sigma'}
\end{equation}
is the effective-mass Hamiltonian with an additional Rashba SO interaction 
term. (We will use $\hbar=1$ units for convenience.)   
This interaction can be interpreted as Zeeman coupling to a $\kk$-dependent
effective magnetic field $2 \lambda (\hat z \times \kk)$. 
The impurity term in the Hamiltonian, 
\begin{equation}
\label{eq:2}
H_i=\int_{\rr} \sum_{\sigma}V_i(\rr) \Psi^{\dag}_{\sigma}(\rr) 
\Psi^{\vphantom \dag}_{\sigma}(\rr)=
\frac{1}{V}\sum_{\kk\kk'\sigma}V_i(\kk-\kk') c^{\dag}_{\kk \sigma}
c^{\vphantom \dag}_{\kk'\sigma},
\end{equation} 
describes the interaction of electrons with an impurity potential 
$V_i(\rr)= u_0 \sum_a \delta(\rr-\rr_a)$.
The spin-independent random potential  
influences the electronic spin state by inducing transitions between momentum 
states that have different Rashba effective fields. 
The SO interaction lifts the spin degeneracy of the effective-mass Hamiltonian
resulting in a momentum-dependent spin-splitting of the conduction band:
\begin{equation}
\label{eq:3}
\epsilon_{\pm}(\kk)=\frac{\kk^2}{2m} \pm \lambda k -\mu.
\end{equation}
We assume here that the Rashba spin-splitting is small compared to the Fermi 
energy $\lambda k_F \ll \epsilon_F$, a good approximation in almost all cases of 
interest.

Our analysis is based on an evaluation of the density-matrix response function 
using standard perturbation-theory methods.\cite{Mahan}
The fundamental object in this approach is the imaginary time Green's 
function  
\begin{equation}
\label{eq:4}
{\cal G}_{\sigma \sigma'}(\rr-\rr',\tau-\tau')=\langle T 
\Psi^{\vphantom \dag}_{\sigma}(\rr,\tau) \Psi^{\dag}_{\sigma'}(\rr',\tau')
\rangle,
\end{equation}
where the angular brackets denote quantum, thermal and disorder averages.
We compute the disorder averaged Green's function in the first Born 
approximation, which implies a self-energy in the Matsubara frequency 
representation given by 
\begin{equation}
\label{eq:7}
\Sigma_{\sigma\sigma'}(i\omega)=-\gamma\int \frac{d^2 k}{(2\pi)^2}
{\cal G}^0_{\sigma \sigma'}(\kk,i\omega),
\end{equation}
where $\gamma=n_i u_0^2$, $n_i$ is the density of impurities and 
${\cal G_{\pm}}^0$ is the Green's function of the clean system without 
impurities. 
The self-energy turns out to be momentum-independent and 
upon analytic continuation, $i\omega \rightarrow \omega+i\eta$, we obtain 
the familiar expression for the Born-approximation retarded self-energy: 
\begin{equation}
\label{eq:8}
\Sigma_{\sigma \sigma'}(\omega+i\eta)=
-\frac{i}{2\tau} \delta_{\sigma \sigma'},
\end{equation}  
where $\tau=1/\pi\gamma\varrho_0$ is the mean scattering time 
and $\varrho_0=m/\pi$ is the total density of states at the Fermi energy.

It is convenient to decompose the disorder-averaged retarded and 
advanced real-time Green's functions $G^{R,A}_{\sigma\sigma'}$ into 
spin-independent {\it singlet} and spin-dependent {\it triplet} parts:
\begin{equation}
\label{eq:9}
G^{R,A}_{\sigma\sigma'}(\kk,\omega)=G^{R,A}_s(\kk,\omega)\delta_{\sigma\sigma'}
+\GG^{R,A}_t(\kk,\omega)\cdot \boldsymbol{\tau}_{\sigma\sigma'},
\end{equation}
where $\tau^a$ are the usual spin-$\frac{1}{2}$ Pauli matrices. 
The singlet and triplet Green's functions are given by 
\begin{eqnarray}
\label{eq:10}
&&G^{R,A}_s(\kk,\omega)=\nonumber \\
&&\frac{1}{2}\left[\frac{1}{\omega-\xi_{\kk}-\lambda k
\pm \frac{i}{2\tau}}+\frac{1}{\omega-\xi_{\kk}+
\lambda k \pm \frac{i}{2\tau}}\right],
\end{eqnarray} 
and 
\begin{eqnarray}
\label{eq:10a}
&&\GG^{R,A}_t(\kk,\omega)=\frac{\hat k \times \hat z}{2} \nonumber \\
&&\times\left[\frac{1}{\omega-\xi_{\kk}-\lambda k
\pm \frac{i}{2\tau}}-\frac{1}{\omega-\xi_{\kk}+
\lambda k \pm \frac{i}{2\tau}}\right], \nonumber \\
\end{eqnarray}
where $\xi_{\kk}=\kk^2/2m - \mu$.

We can now proceed to evaluate the coupled spin and charge density 
response functions.
We introduce the generalized density operator
\begin{equation}
\label{eq:11}
\hat \varrho_{\sigma_1\sigma_2}(\rr,t)=\Psi^{\dag}_{\sigma_2}(\rr,t)
\Psi^{\vphantom \dag}_{\sigma_1}(\rr,t),
\end{equation}
whose expectation value is the density matrix.  (It is the matrix character
of this quantity in spin-space that allows us to look at coupled spin-charge
response; for present purposes it is adequate to
specialize to diagonal elements in position-space.) 
From standard linear-response theory the retarded density response function is given by:
\begin{eqnarray}
\label{eq:12}
&&\chi_{\sigma_1\sigma_2,\sigma_3\sigma_4}(\rr-\rr',t-t')= \nonumber \\
&&- i \theta(t-t')\langle[\hat \varrho^{\dag}_{\sigma_1\sigma_2}
(\rr,t),\hat \varrho^{\vphantom \dag}_{\sigma_3\sigma_4}(\rr',t')]\rangle.
\end{eqnarray}
It is well known \cite{Mahan} that this quantity can be evaluated to leading order
in $1/k_F \ell$ by summing all 
Born approximation self-energy and ladder vertex corrections to the 
polarization bubble (here $\ell=v_F \tau$ is the mean-free-path).  
For $\delta$-function impurities, the 
ladder sum for the Fourier-transformed retarded response function is a 
matrix geometric series which is easy to evaluate.  We find that 
\begin{eqnarray}
\label{eq:13}
&&\chi_{\sigma_1\sigma_2,\sigma_3\sigma_4}(\qq,\Omega)= \nonumber \\
&&-\frac{i\Omega \tau \varrho_0}{2}
I_{\sigma_1\sigma_2,\sigma_1'\sigma_2'}(\qq,\Omega)
{\cal D}_{\sigma_1'\sigma_2',\sigma_3\sigma_4}(\qq,\Omega) \nonumber \\
&&-\frac{1}{2}\varrho_0\delta_{\sigma_1\sigma_3}
\delta_{\sigma_2\sigma_4},
\end{eqnarray}
where
\begin{eqnarray}
\label{eq:14}
&&I_{\sigma_1\sigma_2,\sigma_3\sigma_4}(\qq,\Omega)= \nonumber \\
&&\gamma \int\frac{d^2 k}{(2\pi)^2} G^A_{\sigma_3\sigma_1}(\kk,0)
G^R_{\sigma_2\sigma_4}(\kk+\qq,\Omega),
\end{eqnarray}
and
\begin{equation}
\label{eq:15}
{\cal D}=[1-I]^{-1}
\end{equation}
is the coupled spin-charge diffusion propagator or ``diffuson''. Summation over repeating spin 
indices is implied.

At this point it is convenient to go to a more physical charge-spin-component
representation for the response function:
\begin{equation}
\label{eq:17}
\chi_{\alpha\beta}(\qq,\Omega)=\frac{1}{2}\tau^{\alpha}_{\sigma_1\sigma_2}
\chi_{\sigma_1\sigma_2,\sigma_3\sigma_4}(\qq,\Omega)
\tau^{\beta}_{\sigma_4\sigma_3},
\end{equation}
where $\alpha,\beta=c,x,y,z$.
Inserting the identity matrix resolution 
$\frac{1}{2}\tau^{\alpha}_{\sigma_1\sigma_2}\tau^{\alpha}_{\sigma_2'\sigma_1'}$
between factors in Eq.(\ref{eq:13}), we obtain
\begin{equation}
\label{eq:18}
\chi_{\alpha\beta}(\qq,\Omega)=-\frac{i\Omega\tau\varrho_0}{2}
I_{\alpha\gamma}(\qq,\Omega){\cal D}_{\gamma\beta}(\qq,\Omega)
-\frac{1}{2}\varrho_0\delta_{\alpha \beta}.
\end{equation} 
The integral over momentum in Eq.(\ref{eq:14}) is elementary but 
leads to cumbersome expressions for $I(\qq,\Omega)$ that are listed in the 
appendix. 
We focus on the long-wavelength, low-frequency limit of 
the response function (\ref{eq:18}) in the remaining sections of this paper. 

\section{Spin Transport Equations}
\label{sec:3}  
We are interested in the coupled 
dynamics of spin and charge, coarse-grained over lengths long compared to the 
mean-free path $\ell$ and times long compared to the scattering time $\tau$. 
We concentrate here on the limit of weak SO interactions,
$\lambda k_F \tau \ll 1$ in which the scattering time is much shorter than the 
spin-precession period and the spin-split Rashba bands are therefore not established. 
(The low-frequency, long-wavelength expansion of $I(\qq,\Omega)$ 
is not analytic in the strong SO scattering limit.)
In this diffusive limit, the inverse density fluctuation propagator 
(the diffuson) ${\cal D}^{-1}(\qq,\Omega)=1-I(\qq,\Omega)$ simplifies to:
\begin{eqnarray}
\label{eq:19}
&&{\cal D}^{-1}(\qq,\Omega)= \left(-i\Omega + D \qq^2\right) {\bf 1} + 
\nonumber \\
&&\left(
\begin{array}{cccc} 
0& i \Gamma_{sc} q_y & -i \Gamma_{sc} q_x & 0 \\
i \Gamma_{sc} q_y & 1/\tau_{\perp} & 0 & -i \Gamma_{ss} q_x \\ 
-i \Gamma_{sc} q_x & 0 & 1/\tau_{\perp} & -i \Gamma_{ss} q_y \\
0 & i \Gamma_{ss} q_x & i \Gamma_{ss} q_y & 1/\tau_{z}
\end{array}
\right),\nonumber \\
\end{eqnarray}
where $D=v_F^2 \tau/2$ is the diffusion constant, 
$\tau_{\perp} = 2 \tau / (2 \lambda k_F \tau)^2$ and 
$\tau_z = \tau_{\perp}/2$ are the in plane and out of plane spin relaxation 
times, and $\Gamma_{sc}=-2 \lambda (\lambda k_F \tau)^2$ and 
$\Gamma_{ss}=4\lambda \epsilon_F \tau$
are the spin-charge and in-plane to out-of-plane spin couplings that result
from the SO interactions.  Note that $4 D/\tau_{\perp} \Gamma_{ss}^2 =1$.

Transforming this diffusion propagator to real space and time leads to the 
following system of coupled spin and charge transport equations that is the 
principal result of this paper:
\begin{eqnarray}
\label{eq:20}
\frac{\partial N}{\partial t}&=& D{\boldsymbol \nabla}^2 (N + \varrho_0 V_c)
+ 2 \Gamma_{sc} (\hat{z} \times {\boldsymbol \nabla})
 \cdot ({\bf S} - \varrho_0 {\bf h}) + I^{c}, \nonumber \\
\frac{\partial S^a}{\partial t}&=&\left(D{\boldsymbol \nabla}^2 -
\frac{1}{\tau_a}\right)
(S^a -\varrho_0 h^a) \nonumber \\
&+&\Gamma_{ss} \left[(\hat{z} \times {\boldsymbol \nabla}) \times ({\bf S}
-\varrho_0 {\bf h} ) \right]_a \nonumber \\
&+&\frac{\Gamma_{sc}}{2} (\hat{z} \times {\boldsymbol \nabla})_a 
(N + \varrho_0 V_c) + I^{s,a}.
\nonumber \\
\end{eqnarray}       
In these equations $V_c$ and ${\bf h}$ are the charge and spin (Zeeman) components of 
the external potential.  The last term on the right-hand side of each equation
has been inserted by hand to represent charge and spin currents, 
$I^{c}$ and $I^{s,a}$, vertically injected into the 2DEG. 
The factors of $2$ and $1/2$, that appear in front of the coefficient 
$\Gamma_{sc}$, follow from the relationships between spin and charge densities
and the corresponding combinations of elements of the density matrix. 
Note that in a generalization of the familiar Einstein relations,
the external charge and spin potentials and the 
corresponding chemical potentials, $N/\varrho_0$ and 
${\bf S}/\varrho_0$, are always summed; the charge and spin-densities 
respond as usual to electrochemical potentials and their  
gradients. (A 2DEG system with excess spin and charge densities $N$ and 
${\bf S}$, has excess chemical potential    
$(N \pm 2|{\bf S}|)/\varrho_0$ for spins oriented along and in opposition to 
$\hat{S}$ respectively.) 

A physical understanding of the numerical values and the parametric 
dependences of the coefficients that appear in front of the various terms in 
Eqs.(\ref{eq:20}) is most easily obtained by considering the 
limit in which external potentials are absent.  Then the drift and 
diffusion of charge and spin can be understood by considering the time
evolution of electrons that start at the origin in specified spin-states and 
are scattered randomly between various Rashba states at arbitrary angles on the 
Fermi circle.  These electrons undertake random walks that make correlated
steps of size $\sim \lambda k_F \tau$ in spin-space and $\ell$ in position-space.
The joint probability distribution function that results from these correlated 
changes in spin and position is readily evaluated.  Associating the coarse-grained
spin and charge distributions with the distribution of starting positions and 
spin orientations, the coefficients of n'th derivative terms in 
Eqs.(\ref{eq:20}) arise from 
n'th order spatial moments of the spin and charge diffusion clouds.  For example the 
diffusion constant $D$ is related, as usual, to the second spatial moment of charge diffusion 
cloud and is therefore proportional to the square of the 
spatial step length $\ell$ times the step rate $\tau^{-1}$.  Similarly $\Gamma_{ss}$
is due to spin-precession and is proportional to the first spatial moment of the 
$S^x$ spin projection in the diffusion cloud generated by spins that start with an 
orientation out of the plane.  It is therefore 
proportional to the product of the spin-space and orbital-space step lengths and 
to the step rate.  All non-standard coefficients in our equations can 
be understood in terms of the correlation between velocity and 
spin-precession axis that 
exists throughout the random walk.  This line of argument can be followed to 
provide an independent confirmation of Eqs.(\ref{eq:20}).

\section{Operator Equations of Motion} 
\label{sec:4}
Some insight into our general equations for the diffusive charge and spin 
density dynamics of 2DEG's with the Rashba spin-orbit coupling can be obtained 
by comparing Eqs.(\ref{eq:20}) with the equations of motion of the charge and 
spin-density operators for this system. 
Let us first consider the Heisenberg equation of motion for the charge 
(or, more precisely, particle number) density operator
\begin{equation}  
\label{eq:21}
N(\rr) = \Psi^{\dag}_{\sigma}(\rr)\Psi^{\vphantom \dag}_{\sigma}(\rr).
\end{equation}
The equation of motion reads
\begin{equation}
\label{eq:22}
\frac{\partial N}{\partial t} = i \left[ H , N \right].
\end{equation}
Since the particle number is conserved, we expect this equation to have 
the form of a continuity equation:
\begin{equation}
\label{eq:23}
\frac{\partial N}{\partial t} = - {\boldsymbol \nabla} \cdot {\bf J}^c,
\end{equation}
where ${\bf J}^c$ is the charge current density.  
Fourier transforming the charge density operator and evaluating 
the elementary commutator in Eq.(\ref{eq:22}) implies 
the following expression for the charge current density:
\begin{equation}
\label{eq:24}
{\bf J}^c = -\frac{i}{2m} \left( \Psi^{\dag}_{\sigma} 
{\boldsymbol \nabla} \Psi^{\vphantom \dag}_{\sigma} - h.c. \right)+ 
2 \lambda \left( S^x \hat y - S^y \hat x \right),
\end{equation}
where 
\begin{equation}
\label{eq:25}
S^a(\rr) = \frac{1}{2} \Psi^{\dag}_{\sigma}(\rr) \tau^a_{\sigma \sigma'}
\Psi^{\vphantom \dag}_{\sigma'}(\rr),
\end{equation}
is the $a$-component of the spin density. 
The first term in Eq.(\ref{eq:24}) is the usual quantum-mechanical 
expression for the particle current density.
We will call this contribution to the charge current a {\it kinetic} 
contribution.
As seen from Eq.(\ref{eq:24}), SO interactions result in an additional 
contribution to the charge current density, that we accordingly 
refer to as the {\it spin-orbit} contribution. 
This contribution is proportional to the in-plane spin densities. 
Comparing Eqs.(\ref{eq:23}) and (\ref{eq:24}) with the first of 
Eqs.(\ref{eq:20}), we conclude that the kinetic contribution to the 
charge current transforms in the diffusive limit to a kinetic 
contribution of the standard form, proportional to both 
the diffusion constant and the electrochemical potential gradient.
The SO contribution apparently remains separate and proportional to the
spin-density, rather than being subsumed at long-wavelengths in the 
diffusive term.  The total charge current is therefore given by, 
\begin{eqnarray}
\label{eq:26}
{\bf J}^c&=&-D {\boldsymbol \nabla} (N + \varrho_0 V_c)\nonumber \\ 
&+&2 \Gamma_{sc} \left[ (S^x - \varrho_0 h^x) \hat y - (S^y - \varrho_0 h^y) 
\hat x \right].
\end{eqnarray} 

Let us now repeat the same analysis for the spin density operators. 
Since the spin is not conserved, there is some freedom of choice in how
the spin current density operator is defined. \cite{Rashba03}  
We choose to define the spin current as the symmetrized product 
of the charge current discussed above and the spin operator, a definition 
that seems natural from a microscopic point of view and has been used 
previously, for example in discussing the spin-Hall effect. \cite{Sinova03}  
The spin current density operator is, therefore, also a sum of a kinetic 
and a spin-orbit contribution and has the following form:
\begin{eqnarray}
\label{eq:27}
{\bf J}^{s,x}&=&{\bf J}^{s,x}_{kin} + \frac{\lambda}{2} N \hat y, \nonumber \\
{\bf J}^{s,y}&=&{\bf J}^{s,y}_{kin} - \frac{\lambda}{2} N \hat x, \nonumber \\
{\bf J}^{s,z}&=&{\bf J}^{s,z}_{kin},
\end{eqnarray}
where
\begin{equation}
\label{eq:28}
{\bf J}^{s,a}_{kin} =   
-\frac{i}{4m} \left( \Psi^{\dag}_{\sigma}  
{\boldsymbol \nabla} \Psi^{\vphantom \dag}_{\sigma'} \tau^a_{\sigma \sigma'} 
- h.c. \right),
\end{equation}
is the kinetic contribution to the spin current. 
Note that the current of the $z$-component of the spin has only a kinetic 
component.
The Heisenberg equations of motion for the spin density operators can then 
be written in the following form:
\begin{equation}
\label{eq:29}
\frac{\partial S^a}{\partial t} = - {\boldsymbol \nabla} \cdot 
{\bf J}^{s,a} + F_a,
\end{equation}
where $F_a$ is an additional source term that is given by:
\begin{eqnarray}
\label{eq:30}
F_{x,y}&=&-2 \lambda m J^{s,z}_{x,y}, \nonumber \\
F_z&=&2 \lambda m \left( J^{s,x}_x + J^{s,y}_y \right).
\end{eqnarray}
As before, comparing Eqs.(\ref{eq:27})--(\ref{eq:30}) with Eq.(\ref{eq:20}),
we conclude that the kinetic contribution to the spin currents is proportional
to the gradient of the spin electrochemical potential and, 
in addition, that the currents of the in-plane spin components have SO 
contributions as in the microscopic equations of motion:
\begin{eqnarray}
\label{eq:31}
{\bf J}^{s,x}&=&-D {\boldsymbol \nabla} \left(S^x - \varrho_0 h^x\right)
+ \frac{\Gamma_{sc}}{2} (N + \varrho_0 V_c) \hat y, \nonumber \\
{\bf J}^{s,y}&=&-D {\boldsymbol \nabla} \left(S^y - \varrho_0 h^y\right)
- \frac{\Gamma_{sc}}{2} (N + \varrho_0 V_c) \hat x, \nonumber \\
{\bf J}^{s,z}&=&-D {\boldsymbol \nabla} \left(S^z - \varrho_0 h^z\right).
\end{eqnarray}
As in the charge current case, the spin-orbit contribution to the 
microscopic spin current 
is not subsumed in the diffusive contribution, but appears separately.  
Interestingly, a change occurs in passing from the microscopic expression to the coarse-grained
transport theory expression in that the chemical potential (proportional to the density $N$ in 
2D) is replaced by the electrochemial potential.  Because the spin-orbit 
spin current contributions are proportional
to the charge density in the absence of external fields, they are non-zero in equilibrium, as noted 
by Rashba.\cite{Rashba03}  Although the constant equilibrium spin currents 
in a uniform system have no 
physical consequences as far as we are aware, these spin-orbit terms in the 
spin current do play an important role in coupled spin-charge transport as we 
illustrate in the following section.
Also note that the source terms $F_a$ in the 
microscopic equations of motion appear in an almost, but not completely, 
identical way in the diffusive equations of motions;    
the constants multiplying the currents are 
twice as large in the diffusive case: $\Gamma_{ss}/D = 4 \lambda m$ instead of
$2 \lambda m$ in Eq.(\ref{eq:30}) and $N$ is replaced by its electrochemical 
equivalent $N+\varrho_0 V_c$.    

Finally, let us comment on the relationship between our results and the 
recently 
discovered spin-Hall effect,\cite{Sinova03,Zhang03,Schliemann03,Hu03,Shen03,Shen04,Xie04,Inoue04,Dimitrova04,Mishchenko04,Nomura04} i.e. a transverse 
${\bf J}^{s,z}$ current in response to an in-plane electric field. 
Note that, as in the microscopic expressions,     
the current of the $z$-component of the spin has only a kinetic contribution.
This means that, apparently, the spin-Hall effect 
{\it does not occur in the diffusive limit}, since 
the current of $S^z$ in Eq.(\ref{eq:20}) has only a diffusive contribution, 
that does not react to the electric field. 
In the diffusive regime, the spin current divergence
(from the spin-orbit contribution), produced by a uniform electric field, 
is balanced 
in the steady-state by spin relaxation.  This balancing leads to a 
spin-polarization 
perpendicular to the electric field direction, as noted some time ago.\cite{Levitov85,Magarill01} 
In the regime of resolved spin-orbit induced spin-splitting,
the diffusive transport picture of Eq.(\ref{eq:20}) is no longer applicable.
In this case, the current of the $z$-component of the spin will have a 
contribution, proportional to $\hat z \times {\bf E}$, due to the intrinsic 
spin-Hall effect.    
\section{Applications of Coupled Spin-Charge Transport Equations} 
\label{sec:5} 
 
In the case of an infinite 2DEG , Eqs.(\ref{eq:20}) 
can be solved by Fourier transformation. 
Rotating coordinate axes so that the $y$-axis is along the direction of 
$\qq$, brings the inverse diffusion propagator Eq.(\ref{eq:19}) to 
the block-diagonal form:
\begin{eqnarray}
\label{eq:32}
&&{\cal D}^{-1}(\qq,\Omega)= \left(-i\Omega + D \qq^2\right) {\bf 1} + 
\nonumber \\
&&\left(
\begin{array}{cccc} 
0& i \Gamma_{sc} q & 0 & 0 \\
i \Gamma_{sc} q & 1/\tau_{\perp} & 0 & 0 \\ 
0 & 0 & 1/\tau_{\perp} & -i \Gamma_{ss} q \\
0 & 0 & i \Gamma_{ss} q & 1/\tau_{z}
\end{array}
\right),\nonumber \\
\end{eqnarray}
The eigenmodes are then easily calculated to be:
\begin{eqnarray}
\label{eq:33}
i\Omega_{1\pm}&=&D\qq^2 +\frac{1}{2\tau_{\perp}} \pm 
\sqrt{\frac{1}{4\tau_{\perp}^2} - \Gamma_{sc}^2 \qq^2}, \nonumber \\
i\Omega_{2\pm}&=&D\qq^2 + \frac{\tau_{\perp}+\tau_z}{2\tau_{\perp}\tau_z}
\pm \sqrt{\left(\frac{\tau_{\perp}-\tau_z}{2\tau_{\perp}\tau_z}\right)^2 +
\Gamma_{ss}^2 \qq^2}. \nonumber \\
\end{eqnarray}
The $i\Omega_{1\pm}$ modes correspond to coupled diffusion of charge and
the in-plane spin density component that is transverse to the direction
of $\qq$, i.e. $S^x$ in this convention.  
Note, that the mode $i\Omega_{1-}$ is gapless at $\qq=0$.
This means that this mode corresponds to a (nearly) conserved quantity,
with a very long relaxation time at small $\qq$. 
Exactly at $\qq=0$ this quantity is of course simply the conserved total 
particle number.   
However, at finite wavevectors it corresponds to a 
linear combination of the charge density and the $x$-component of the spin 
density.

The $i\Omega_{2\pm}$ modes correspond to coupled diffusion of 
$S^y$ and $S^z$ spin densities.
This coupling originates from the Rashba spin precession as explained above. 
Note that $i\Omega_{2-}$ has a minimum at a finite wavevector
$q^*=\sqrt{15}\lambda m/2$, as discovered previously in 
Ref.[\onlinecite{Froltsov01}]. 
This means that the $S^{y,z}$ Fourier component with the slowest relaxation rate 
will actually be at $q=q^*$, unlike in the case of the ordinary diffusive 
relaxation, where the slowest relaxation rate is at $q=0$.  

\begin{figure}[t]
\includegraphics[width=8cm]{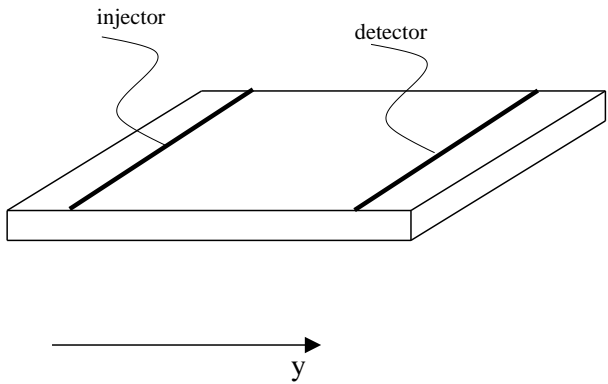}
\caption{Cartoon of the spin injection experiment.
A polarized current, that has a polarization component along the $x$-direction
(i.e. along the electrode) is injected at the left electrode and 
collected at the right one. A voltage develops between the electrodes that 
depends on the spin-polarizations of both emitter and collector.} 
\label{fig:1}
\end{figure}

Let us now look at stationary solutions of Eq.(\ref{eq:20}). 
For simplicity and clarity of presentation we will assume that spin 
and charge densities are uniform in the $x$-direction, and, therefore, 
the inverse diffusion propagator has the simple block-diagonal form 
Eq.(\ref{eq:32}).
In the following paragraphs we discuss the stationary state response of the 2DEG system to external 
spin and charge currents injected or drained along lines of constant $y$ as illustrated in Fig. \ref{fig:1}.
We first consider the response to a flux of the $z$-component of the 
spin, $I^{s,z}$. 
Inverting the lower block of the inverse diffusion propagator Eq.(\ref{eq:32}),
we obtain:
\begin{equation}
\label{eq:34}
S^z(\qq) = I^{s,z} \frac{D\qq^2 + 1/\tau_{\perp}}{D^2 \qq^4 - 
4(\lambda \epsilon_F
\tau)^2 \qq^2 + 32 (\lambda^2 m \epsilon_F \tau)^2},
\end{equation} 
and 
\begin{equation}
\label{eq:35}
S^y(\qq) = I^{s,z} \frac{-i \Gamma_{ss} q}{D^2 \qq^4 - 
4(\lambda \epsilon_F
\tau)^2 \qq^2 + 32 (\lambda^2 m \epsilon_F \tau)^2},
\end{equation} 
The Fourier transform to real space is readily evaluated by contour
integration.  Poles occur at the roots of the denominator located at 
\begin{equation}
\label{eq:36}
q = \pm \lambda m \sqrt{2 (1 \pm i \sqrt 7)}.
\end{equation}
Note that all the roots are complex. This means that the nonequilibrium spin 
density profile in 
this case will not have the usual form, exponentially 
decaying away from the point where the current is injected, with a 
characteristic spin diffusion decay length.  (Because the 
spin-orbit coupling is linear in momentum, the distance travelled by a 
Fermi energy electron during one spin-precession, $\propto 1/\lambda m$, 
is independent of the Fermi momentum.)  
Instead it will clearly involve an oscillatory component, which 
is the remnant of the Rashba spin precession in the diffusive regime. 
The inverse of the characteristic decay length 
and the inverse period of the spatial oscillations are given by the imaginary
and real parts of $\lambda m \sqrt{2(1 + i \sqrt7)}$ correspondingly.

Let us now turn to the more interesting issue of signatures 
of the spin-charge coupling in our transport equations in
spin injection experiments.
We imagine the geometry schematically depicted
in Fig. \ref{fig:1}.
Assume two infinitely long ferromagnetic electrodes are placed on top of the 
2DEG sample a distance $L$ from each other.
Let a charge current $I$, polarized in the $x$-direction (i.e. along the 
electrode), be injected into the 2DEG from the $y=0$ electrode.
Assume that the degree of spin polarization of this current is $\alpha$,
i.e. the injected spin current is $I^{s,x}=\alpha I$.
Assume in addition that this current is extracted at the second electrode at 
$y=L$, 
which has a degree and sign of spin polarization denoted by $\beta$, that can
differ from $\alpha$.  
This circumstance is expressed compactly by the following source terms in our 
spin-charge transport equations:
\begin{eqnarray}
\label{eq:37}
I^c(y)&=&I \left[ \delta(y) - \delta(y-L) \right], \nonumber \\
I^{s,x}(y)&=&I\left[\alpha \delta(y) - \beta \delta(y-L) \right].
\end{eqnarray}
We now evaluate voltage, i.e. the electrochemical potential difference,
that develops between the two electrodes and discuss how it depends on  
the relative spin polarization of the electrodes.
    
Inverting the upper block of the diffusion propagator Eq.(\ref{eq:32}), we
obtain the Fourier transformed local electrochemical potential change  
$U= N/\varrho_0+V_c$ that is generated
in the 2DEG in response to the injected spin-polarized current:
\begin{eqnarray}
\label{eq:38}
\varrho_0 U(q) &=&I\left(1 - e^{-i q L} \right) \frac{1}{D q^2} \nonumber \\
&-&I\left( \alpha - \beta e^{-i q L} \right) \frac{2 i \Gamma_{sc}}
{D^2 q \left(q^2 + 1/D\tau_{\perp}\right)}.
\end{eqnarray}   
We define the effective electric field in the 2DEG in the usual way 
in terms of the gradient of the electrochemical potential:
\begin{equation}
\label{eq:39}
{\bf E} = - \frac{1}{e} \boldsymbol{\nabla} U.
\end{equation}
The electric field response to the injected current can then be easily 
calculated by an inverse Fourier transformation:
\begin{eqnarray}
\label{eq:40}
&&E(y)=-\frac{i}{2 \pi e} \int_{-\infty}^{\infty} dq \, q \,U(q) 
e^{i q y} = \frac{I}{e \varrho_0 D} \nonumber \\ 
&\times&\left[1 + \Gamma_{sc}\sqrt{\tau_{\perp}/D}
\left(\alpha e^{-y/\sqrt{D \tau_{\perp}}} - \beta e^{-(L-y)/
\sqrt{D \tau_{\perp}}}\right)\right]. \nonumber \\
\end{eqnarray}
The voltage between the electrodes is therefore given by:
\begin{eqnarray}
\label{eq:41}
&&V=\int_0^L E(y) dy \nonumber \\
&=&\frac{I L}{e \varrho_0 D} \left[ 1 + 
\frac{\Gamma_{sc} \tau_{\perp}}{L}\left(1 - e^{-L/\sqrt{D \tau_{\perp}}}
\right) (\alpha - \beta)\right].
\end{eqnarray}
Note that in addition to the usual Ohm's law contribution to the 
potential drop, the first term on the right hand side of Eq.(\ref{eq:41}), 
there is also a contribution  
proportional to the difference of spin polarizations of the two 
ferromagnetic electrodes.  
This sensitivity of the resistance of a paramagnetic system to the spin-polarization of 
the current-carrying electrons results from the spin-charge coupling terms in our transport
equations that generate a contribution to the electric field 
proportional to $\Gamma_{sc}$ (see Eq.(\ref{eq:40})). 
Note that this contribution to the voltage is present 
even when the electrodes are separated by a distance larger than the 
spin-coherence length. 

This effect could be studied by attaching voltage probes to the 2DEG near the 
ferromagnetic electrodes, or simply by measuring the voltage drop
between the ferromagnetic electrodes.  In the latter case, the total
voltage will contain contact contributions from the local voltage drops
between the ferromagnetic electrodes and the 2DEG.  The spin-orientation
dependent voltages that we discuss  will, in general, need to 
be distinguished from other spin polarization-dependent voltages that 
occur in magnetotransport, for example the spin-polarization dependent 
open-circuit voltages measured by ferromagnetic electrodes first discovered
in the seminal work of Johnson and Silsbee. \cite{Silsbee85} 
The effects that we discuss here can be distinguished in several ways.
First of all, the voltage differences that we have calculated are ones 
that would be measured by paramagnetic voltage probes.  Secondly,
our voltages have a characteristic dependence on the polarization of the 
injected spin current. 
In the case of the effect described by Eq.(\ref{eq:41}), the voltage drop 
is maximal when the first electrode is polarized along the
$x$-direction, while the second one is polarized along the $-x$-direction.
On the other hand, if the electrodes are polarized along the $y$ and 
$-y$-directions, the polarization-dependent voltage drop will vanish. 

\section{Conclusions}
\label{sec:6}
In this paper we have examined the issue of the diffusive spin and charge 
density transport in Rashba 2DEG systems.
The separation between the spin and momentum relaxation time scales in the 
diffusive regime has allowed us to use a statistical description, where 
the spin and charge transport is described by local spin and charge 
electrochemical potentials and their gradients.
Our theory thus generalizes the usual two-component theory of diffusive 
spin transport, \cite{Son87,Fert93} that has found numerous successfull
applications, in particular in the theory of spin-dependent transport in 
magnetic multilayers. \cite{Fert93}     

Our equations with appropriate boundary conditions can be used to model 
experiments on coupled spin-charge transport in 2DEG systems with the Rashba 
SO interactions, involving both electrical and optical spin injection. 
By comparing our equations, valid in the diffusive transport regime,
with the exact operator equations of motion, we have inferred relationships 
between spin and charge current densities and 
spin and charge electrochemical potentials and their gradients. 
These expressions can be used to devise appropriate boundary conditions 
that are necessary to supplement our transport equations at the 2DEG 
boundaries.

As an example of the application of our equations, we have considered 
a simple electrical spin injection experiment and shown that a 
voltage will develop 
between two ferromagnetic contacts if a spin-polarized current is injected 
into a 2DEG sample, that depends on the relative magnetization 
orientation of the 
two contacts.  Unlike the giant magnetoresistance and other familiar 
magnetoresistive effects in spintronics, this voltage drop is present 
even when the distance between the electrodes exceeds the spin-diffusion
length.

\begin{acknowledgments}    

We are grateful for helpful discussions with Ramin Abolfath, Leon Balents, 
Dimi Culcer, Qian Niu, Jairo Sinova and Ulrich Z\"ulicke.
We would like to thank the authors of Ref.[\onlinecite{Mishchenko04}]
for pointing out an error in the calculation of the spin-charge coupling
coefficient in an earlier version of this paper. 
This work was supported by DARPA/ONR N00014-99-1-1096, the Welch Foundation, 
and by the National Science Foundation under grant DMR0115947.   

\end{acknowledgments}

\section{Appendix}
\label{sec:append}
This appendix summarizes some technical details of the 
density matrix response function calculation in 
section \ref{sec:2}.
The main technical problem is the evaluation of the matrix elements 
$I_{\sigma_1\sigma_2,\sigma_3\sigma_4}(\qq,\Omega)$,
which turn out to have the following general form:
\begin{eqnarray}
\label{eq:a1}
&&I_{\sigma_1\sigma_2,\sigma_3,\sigma_4}(\qq,\Omega) = I^{ss}(\qq,\Omega)
\delta_{\sigma_1\sigma_3} \delta_{\sigma_2\sigma_4} \nonumber \\
&+&I^{st}(\qq,\Omega) \delta_{\sigma_1\sigma_3} 
\tau^n_{\sigma_2\sigma_4}
- I^{st *}(\qq,-\Omega) \tau^n_{\sigma_3\sigma_1} \delta_{\sigma_2\sigma_4} 
\nonumber \\ 
&+&\frac{I^{tt}_+(\qq,\Omega)+I^{tt}_-(\qq,\Omega)}{2} 
\tau^q_{\sigma_3 \sigma_1} \tau^q_{\sigma_2\sigma_4} \nonumber \\ 
&+& \frac{I^{tt}_+(\qq,\Omega)-I^{tt}_-(\qq,\Omega)}{2} 
\tau^n_{\sigma_3\sigma_1}\tau^n_{\sigma_2\sigma_4},
\end{eqnarray} 
where 
\begin{eqnarray}
\label{eq:a2}
\tau^q = \left(
\begin{array}{cc}
0&e^{-i \varphi} \\
e^{i\varphi}&0
\end{array}
\right),
\end{eqnarray}
and 
\begin{eqnarray}
\label{eq:a3}
\tau^n = \left(
\begin{array}{cc}
0&-i e^{-i \varphi} \\
i e^{i\varphi}&0
\end{array}
\right),
\end{eqnarray}
are the Pauli matrix components along and perpendicular to the direction 
of $\qq$ ($\varphi$ is the angle between $\qq$ and the $x$-axis).
The explicit expressions for the functions $I^{ss}, I^{st}, I^{tt}_+$ and
$I^{tt}_-$, that appear in the matrix elements of $I$, 
are: 
\begin{equation}
\label{eq:a4}
I^{ss}(\qq,\Omega)=\frac{1}{4}\left[\frac{2}{f_0}
+\frac{1}{f_-}+\frac{1}{f_+}\right],
\end{equation}
\begin{eqnarray}
\label{a5}
&&I^{st}(\qq,\Omega)= \frac{i}{4\sqrt{2D\qq^2\tau}} \nonumber \\
&\times&\left[\sqrt{1-\frac{2D\qq^2\tau}{f_-^2}}-
\sqrt{1-\frac{2D\qq^2\tau}{f_+^2}}\right] - 
i q \lambda \tau (\lambda k_F \tau)^2, \nonumber \\
\end{eqnarray}
\begin{equation}
\label{eq:a6}
I^{tt}_+(\qq,\Omega)=\frac{1}{4}\left[\frac{2}{f_0}
-\frac{1}{f_-}-\frac{1}{f_+}\right],
\end{equation}
\begin{eqnarray}
\label{eq:a7}
&&I^{tt}_-(\qq,\Omega)=\frac{1}{4 D\qq^2\tau} \nonumber \\
&\times&\left[2\frac{f_0^2-D\qq^2\tau}{f_0}
-\frac{f_-^2-D \qq^2 \tau}{f_-}-
\frac{f_+^2-D \qq^2 \tau}{f_+}\right],\nonumber \\
\end{eqnarray}
where
\begin{eqnarray}
\label{eq:a8}
f_0&=&\sqrt{(1-i\Omega\tau)^2+2D\qq^2\tau},\nonumber \\
f_{\pm}&=&\sqrt{(1-i\Omega\tau \pm i\Omega_{so}\tau)^2+2D\qq^2\tau},
\end{eqnarray}
and $\Omega_{so}=2\lambda k_F$ is the Larmor precession frequency associated
with the Rashba field.
The spin-charge coupling is generated by the last term in the expression 
for $I^{st}(\qq,\Omega)$. 
Note that, unlike other elements of the matrix $I(\qq,\Omega)$, 
the term responsible for the spin-charge coupling can only be calculated 
perturbatively in $\lambda$ and $\qq$.   

Expanding Eqs.(\ref{eq:a4})--(\ref{eq:a8}) to leading order in $i\Omega\tau$,
$D\qq^2 \tau$ and $\lambda k_F \tau$, one arrives at Eq.(\ref{eq:20}),
which we have concentrated on in this paper. 
These expressions should be useful in their original form, however, 
in systems with the Rashba interactions strong enough that the 
spin-splitting of the band energies is not smeared out 
by disorder.

\end{document}